# The role of device asymmetries and Schottky barriers on the helicity-dependent photoresponse of 2D phototransistors.

*Jorge Quereda[1,2]\*, Jan Hidding[1], Talieh S. Ghiasi[1], Bart J. van Wees[1],*
*Caspar H. van der Wal[1] and Marcos H.D. Guimaraes[1].*

[1] Zernike Institute for Advanced Materials, University of Groningen, Groningen, The Netherlands.
[2] Nanotechnology Group, USAL-Nanolab, Universidad de Salamanca, Salamanca, Spain.
\* e-mail: j.quereda@usal.es

**Abstract:** Circular photocurrents (CPC), namely circular photogalvanic (CPGE) and photon drag effects, have recently been reported both in monolayer and multilayer transition metal dichalcogenide (TMD) phototransistors. However, the underlying physics for the emergence of these effects are not yet fully understood. In particular, the emergence of CPGE is not compatible with the $D_{3h}$ crystal symmetry of two-dimensional TMDs, and should only be possible if the symmetry of the electronic states is reduced by influences such as an external electric field or mechanical strain. Schottky contacts, nearly ubiquitous in TMD-based transistors, can provide the high electric fields causing a symmetry breaking in the devices. Here, we investigate the effect of these Schottky contacts on the CPC by characterizing the helicity-dependent photoresponse of monolayer $MoSe_2$ devices both with direct metal-$MoSe_2$ Schottky contacts and with h-BN tunnel barriers at the contacts. We find that, when Schottky barriers are present in the device, additional contributions to CPC become allowed, resulting in emergence of CPC for illumination at normal incidence.

## 1. Introduction

Two-dimensional (2D) transition metal dichalcogenides (TMDs) offer a privileged material platform for the realization of ultrathin and efficient optoelectronics.[1,2] Their strong optical absorption, fast optoelectronic response, and high power conversion efficiencies, combined with functional properties such as flexibility, transparency or self-powering make these materials highly promising for the development of novel optoelectronic devices[3–7].

A particularly interesting feature of 2D-TMDs is the coupling between their spin and valley degrees of freedom[8]. In these materials, the optical bandgap is located at two non-equivalent valleys in the reciprocal space, usually labeled as K and K', presenting different optical selection rules and opposite spin-orbit splitting both in the valence band and in the conduction band. In consequence, upon band-edge optical excitation with circularly polarized light, the spin and valley degrees of freedom of the optically excited electrons can be controlled by appropriately selecting the illumination wavelength and helicity[9]. It was recently shown that when a monolayer TMD (1L-TMD) is illuminated at an oblique angle with respect to the crystal plane, a helicity-dependent photocurrent (circular photocurrent, CPC) emerges. This effect has been attributed to circular photogalvanic (CPGE) and photon drag (CPDE) effects [10–12] and opens new exciting possibilities for the realization of 2D self-powered optoelectronic and opto-spintronic devices.

The physical origin of CPCs in 1L-TMDs is still far from understood. In particular, the emergence of CPGE requires a low crystal symmetry not compatible with the $D_{3h}$ symmetry found in pristine 1L-TMDs. Therefore, it requires an external agent – such as mechanical strain or a strong external electric field – to reduce the crystal symmetry to, at most, a single mirror-plane symmetry[11]. One possible agent that can cause this symmetry breaking is the strong electric field that emerges in Schottky contacts to 1L-TMDs[13] when an external bias is applied. In a recent work[11] we studied CPC in a hexagonal boron nitride (h-BN) encapsulated 1L-MoSe$_2$ phototransistor. There, the Schottky barriers were expected to be suppressed, or at least largely reduced, by the presence of bilayer h-BN tunnel barriers between the metallic contacts and the 1L-MoSe$_2$ channel.[14,15] Here, to clarify the role of Schottky barriers we investigate helicity-dependent photocurrents in 1L-MoSe$_2$ devices both with direct metal/MoSe$_2$ contacts and with metal/h-BN/MoSe$_2$ tunnel contacts. In both cases, we observe a CPC that is maximized for an illumination wavelength λ = 790 nm (matching the room-temperature A-exciton resonance of 1L-MoSe$_2$), increases with the gate voltage, and depends non-trivially on the source-drain voltage and the light incidence angle. However, for the direct metal contact geometry, a nonzero drain-source voltage applied between the sensing contacts is needed to obtain a measurable photocurrent, while for the device with h-BN tunnel barriers a nonzero CPC can be clearly observed even at zero drain-source voltage. We find that for devices with direct metal/MoSe$_2$ contacts where asymmetric Schottky barriers are expected to be present, a nonzero CPC emerges even for light incident normal to the crystal plane. This is contrary to the case of the device with h-BN tunnel barriers. Our results thus confirm that the presence of strong, anisotropic electric fields near the direct metal/MoSe$_2$ contacts reduces the symmetry of the MoSe$_2$ channel, leading to the emergence of additional contributions to the CPC, not present in devices with tunnel h-BN contacts.

The contact-dependent contributions to CPC observed here could also be present in earlier reported measurements attributed to the valley-hall effect[16] (VHE) and to a Berry-curvature induced circular photogalvanic effect[10]. These additional contributions to the observed CPC could be distinguished by their characteristic dependence on the illumination angle. Our results thus demonstrate the crucial importance of angle-resolved measurements for an adequate characterization of helicity-dependent optoelectronic effects in 2D systems.

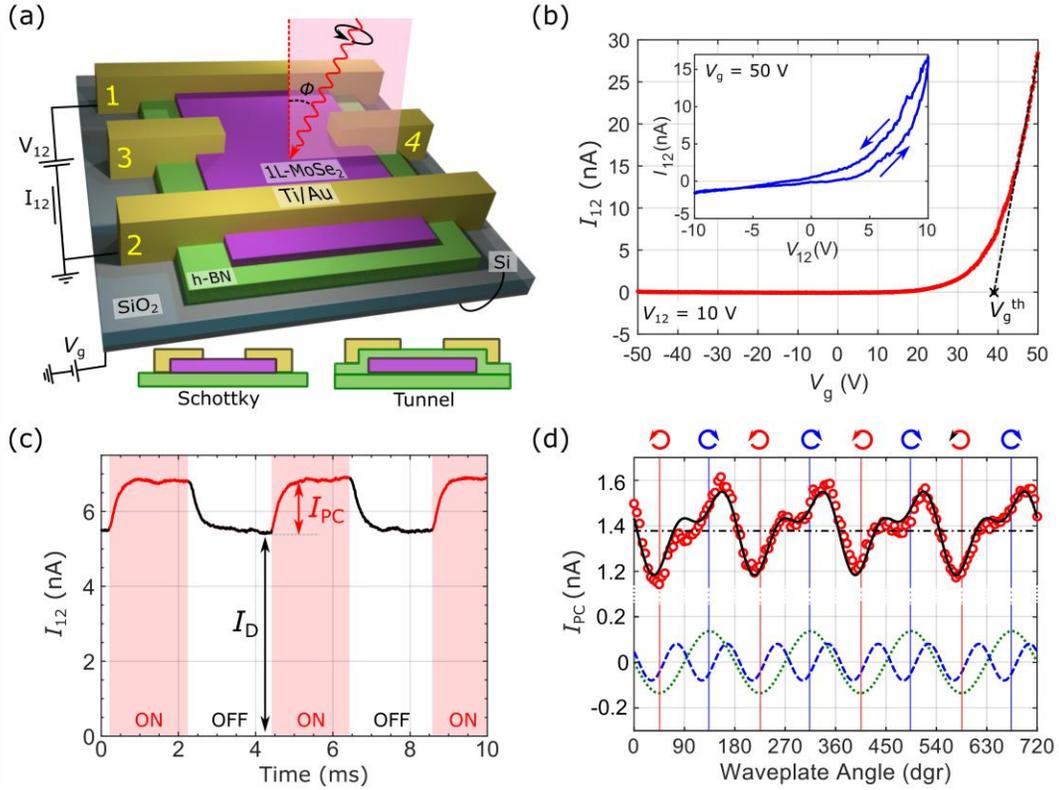

**Figure 1** – Device geometry and optoelectronic response. (a) Schematic of the device with direct metal / 1L-MoSe$_2$ contacts and measurement geometry. The optical excitation is achieved by exposing the entire device to a wavelength-tuneable laser source, hitting the sample at an oblique angle of incidence $\phi$. The polarization and helicity of the light excitation is selected using a $\lambda/4$ waveplate. The bottom diagrams show a side view of the two possible device geometries, either with Ti/Au Schottky contacts fabricated directly on top of the 1L-MoSe$_2$ / hBN structure (left) or with a top bilayer hBN flake acting as tunnel barrier between the 1L-MoSe$_2$ and the contacts (right). (b) Two-terminal transfer characteristic of the device for $V_{12} = 10$ V, showing a clear n-type behaviour. The threshold gate voltage is found to be around $V_g^{th} = 40$ V. Inset: I-V characteristic measured at $V_g = 50$ V. Arrows indicate scan direction. (c) Total current $I_{12}$ along the device for $V_{ds} = 10$ V and $V_g = 50$ V. When the light excitation ($\lambda = 790$ nm) is turned on, the total current along the device increases by $I_{PC} = 1.4$ nA. (d) Photocurrent $I_{PC}$ (red circles) as a function of the waveplate rotary angle $\alpha$ and fitting to eq. 1 (black, solid line). The discontinuous lines represent the three separate contributions indicated in eq. 1, $I_0$ (black, dash-dotted line), $L \sin(4\theta + \delta)$ (blue, dashed line) and $C \sin(2\theta)$ (green, dotted line).

## 2. Device fabrication, electrical characterization and measurement geometry

Figure 1a shows a sketch of the device with direct metal/1L-MoSe$_2$ contacts (a microscope image of the actual device can be found in Supplementary Note 1). We first exfoliated and identified 1L-MoSe$_2$ and multilayer h-BN flakes by standard micromechanical cleavage, and confirmed their flake thickness by atomic force microscopy (see Supplementary Note 1). Then, we used a dry, adhesive-free pick up technique[17] to fabricate the 1L-MoSe$_2$/h-BN heterostructure on a SiO$_2$ (285 nm)/p-doped Si substrate. Finally, we fabricated Ti (5nm)/Au (75 nm) electrodes on top of the structure by standard electron-beam lithography and metal evaporation. We follow a similar fabrication process for the device with hBN tunnel barriers, as detailed in the *methods* section. In that case, the 1L-MoSe$_2$ channel is fully encapsulated between a thick h-BN layer and a bilayer hBN, and metallic electrodes are fabricated directly on top of the bilayer h-BN.

For all the measurements described below, the devices were kept in vacuum and at room temperature. Figure 1b shows a two-terminal transfer characteristic of the non-encapsulated device, presenting a clear n-type behavior with a threshold gate voltage $V_g^{th}$ = 40 V. The *I-V* characteristic (inset in figure 1b) is highly nonlinear, due to the presence of asymmetric Schottky barriers at the metal-MoSe$_2$ contacts. A similar nonlinear *I-V* characteristic is also observed for the device with h-BN tunnel barriers, as discussed in detail in ref. 14.

For characterizing the device photoresponse, we uniformly illuminate the whole sample using a wavelength-tunable continuous wave (CW) laser source and measure the resulting photocurrent. Importantly for our measurements, we use collimated light for optical excitation, as opposed to focusing the light with a high numerical aperture microscope objective. This illumination geometry allows us to control precisely the light incidence angle $\phi$. Figure 1c shows the registered source-drain current $I_{12}$ when the laser source is turned on and off using a chopper for source-drain voltage $V_{12}$ = 10 V, gate voltage $V_g$ = 50 V, illumination wavelength $\lambda = 790$ nm, linear polarization, and light incidence angle $\phi\ =\ 30°$.

When the light is turned on, electrons in the MoSe$_2$ valence band undergo an optical transition to the conduction band, either directly or by formation of excitons, which results in an increase in the conductivity (photoconductivity). Thus, the current flowing through the device increases by $I_{\mathrm{PC}}$. In the measurements discussed below, $I_{\mathrm{PC}}$ is registered using a lock-in amplifier set at the frequency of the mechanical chopper.

In a 2D-TMD phototransistors, photoconductivity can emerge from two main coexisting mechanisms:[18,25–29] photoconductive effect, where light-induced formation of electron–hole pairs leads to an increased charge carrier density; and photovoltaic effect, where light-induced filling or depletion of localized states results in a shift of the Fermi energy. When the characteristic relaxation times for these localized states are very long, photovoltaic effects appear as photodoping, and the Fermi energy shift remains for a long time, or even permanently, after the optical excitation is turned off.[30]

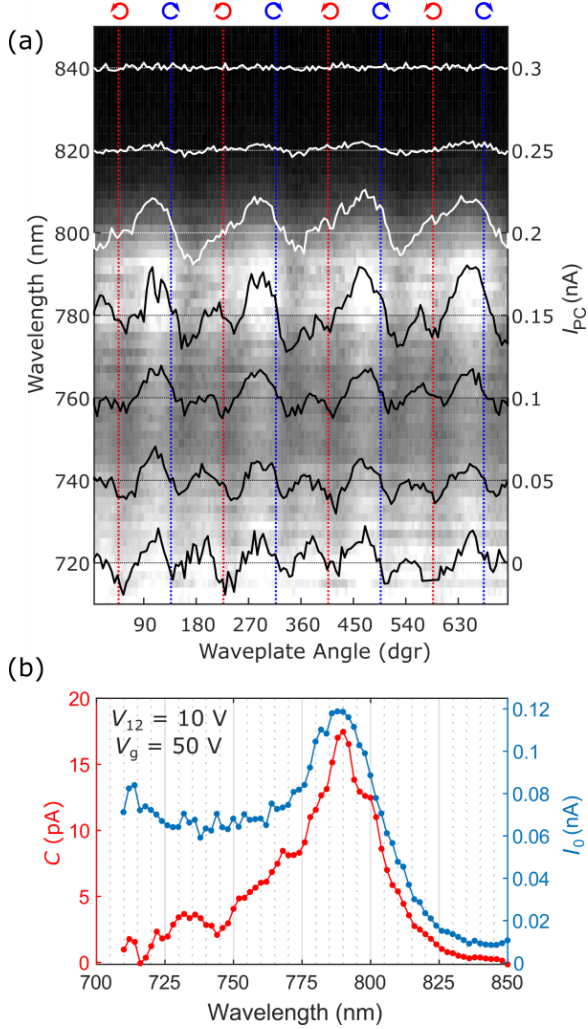

**Figure 2** – Spectral dependence of the $C$ fitting parameter for $\phi = 30°$, $V_{12} = 10$ V and $V_g = 50$ V. (a) Colormap of $I_{PC}$ as a function of the waveplate angle (x axis) and the excitation wavelength (left axis). The solid lines (right axis) show individual $I_{PC}$ profiles at equispaced wavelengths between 720 and 840 nm. For clarity, the base level of these profiles has been shifted vertically in steps of 0.05 nA. (b) $I_0$ (blue, right axis) and $C$ (red, left axis) parameters, obtained from least-square fitting of eq. 1 to the data shown in (a), as a function of the excitation wavelength.

To characterize the helicity-dependent photoresponse of our device we tune the polarization of the incident light by a $\lambda/4$ waveplate. Over a 360° waveplate rotation, the light is modulated twice between left and right circular polarization. Figure 1d shows the helicity-dependent photocurrent $I_{PC}$ as a function of the angle $\theta$ of the fast axis of the waveplate with respect to the polarization axis of the incoming laser. The resulting signal $I_{PC}(\theta)$ can be phenomenologically described as

$$I_{PC}(\theta) = I_0 + C\,sin(2\theta) + L\,sin(4\theta + \delta). \quad (1)$$

Here, $I_0$, $C$ and $L$ respectively account for the polarization-independent, helicity-dependent, and linear polarization-dependent components of $I_{PC}$. Note that the helicity-dependent component $C\,sin(2\theta)$ must be zero for $\theta = 0$ (waveplate fast axis aligned with incident polarization), which corresponds to the output bean being fully linearly polarized. In contrast, the linear polarization-dependent part $L\sin(4\theta + \delta)$ can in principle be maximal for any arbitrary angle, depending on the relative orientation of the incident light polarization and the device. Thus, a phase $\delta$ must be included in the equation. It is also worth remarking that eq. 1 is purely phenomenological, and no assumption is made regarding the microscopic origin of the linear- and helicity-dependent components. In particular, $C$ can include contributions from several effects, including CPGE and CPDE.

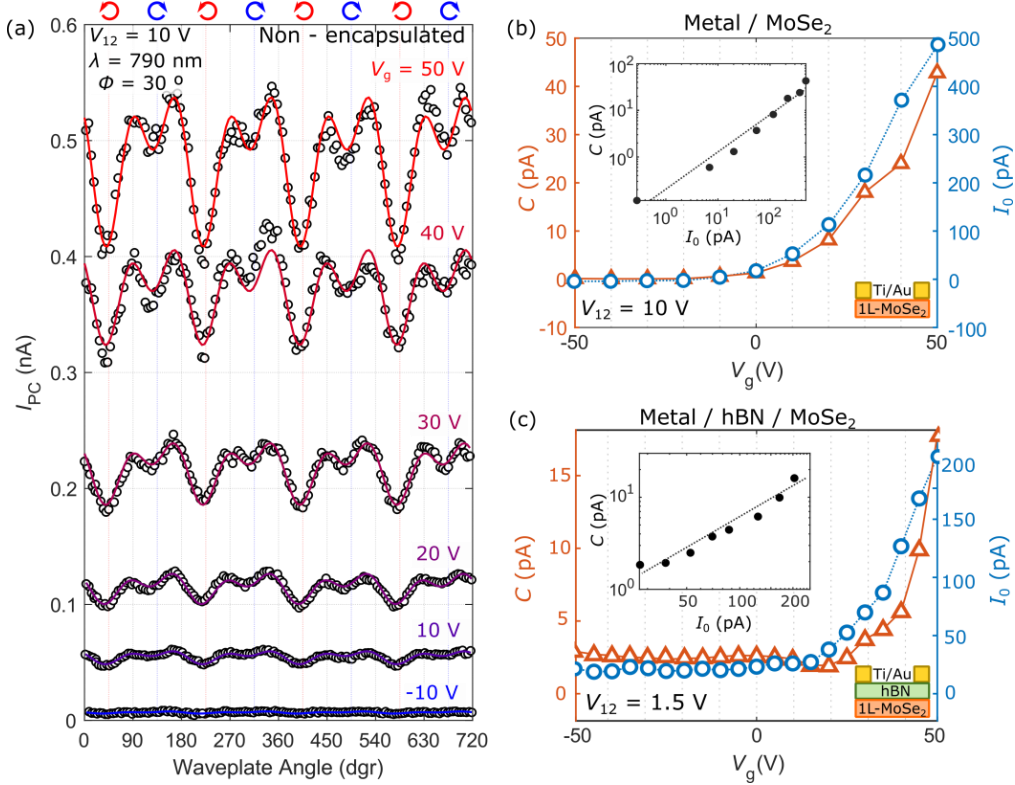

**Figure 3** – Effect of the gate voltage on the helicity-dependent photocurrent. (a) Measured photocurrent $I_{PC}$ in the non-encapsulated device for different gate voltages $V_g$, from -10 V to 50 V, as a function of the $\lambda/4$ waveplate angle. The black solid lines are least-square fits of the experimental data to equation 1. (b-c) $V_g$ dependence of the $I_0$ (blue circles, right axis) and $C$ (orange triangles, left axis) photocurrent components for the non-encapsulated (b) and hBN-encapsulated (c) devices. The insets in pannels (b) and (c) show $C$ as a function of $I_0$ in logarithmic scale.

## 3. Spectral behavior of CPC

Figure 2 shows the spectral dependence of the polarization-independent ($I_0$) and helicity-dependent ($C$) photocurrent components, measured in two-terminal configuration using contacts 1 and 2, with $V_{12} = 10$ V, $V_g = 50$ V, and $\phi = 30°$. Both $I_0$ and $C$ are peaked around $\lambda \approx 790$ nm, matching the wavelength of the 1L-MoSe$_2$ A-exciton resonance [18–20]. For off-resonance wavelengths shorter than 775 nm, $C$ becomes strongly suppressed, even when $I_0$ still remains large. This result is consistent with our earlier measurements in hBN-encapsulated devices[11] and with recent optical measurements showing that light-induced valley population imbalance under off-resonance excitation is rapidly relaxed by intervalley scattering of high-energy excited carriers.[9,21] Therefore, resonant exciton absorption is necessary for efficient CPC generation. For excitation wavelengths longer than $\lambda = 825$ nm only a small polarization-independent photocurrent is observed.

## 4. Dependence of CPC on the gate voltage

Next, we investigate the effect of the gate voltage on the photocurrent. We apply gate voltages between $V_g = -50$ V and $+50$ V while keeping a constant drain-source voltage $V_{12} = 10$ V and illuminating the sample at $\lambda = 790$ nm and $\phi = 30°$. Figure 3a shows the registered photocurrent for the device with direct metal/MoSe$_2$ contacts as a function of the incident light polarization. $I_0$, $C$ and $L$ can be extracted from fittings to equation 1 as described above. Figure 3b shows $I_0$ and $C$ as a function of the gate voltage. Both for $I_0$ and $C$, a nonzero signal can only be observed at gate voltages near to or above $V_{th}$. A similar gate dependence of photoconductivity has been earlier observed in TMD phototransistors,[22,23] and indicates that the observed photoconductivity originates mainly from the photovoltaic effect mentioned in section 2. Thus, the effect of the gate voltage is simply to modulate the overall photoresponse of the device, but does not change the ratio between $I_0$ and $C$. As shown in figures 3b and 3c, the described behavior is observed both for samples with direct metal/MoSe$_2$ contacts and with hBN tunnel barriers. However, for the sample with tunnel contacts, a nonzero $I_0$ and $C$ can be observed even at $V_g < V_{th}$, indicating the presence of an additional contribution to photocurrent. We attribute this new contribution to an enhanced photoconductive effect in hBN-encapsulated samples[22].

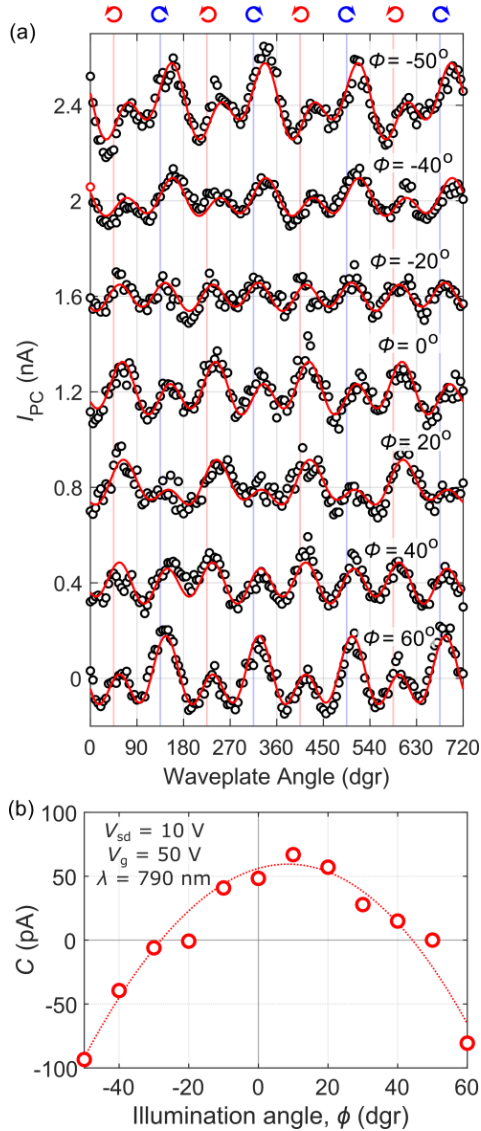

**Figure 4** – Dependence of $C$ on the illumination angle. (a) Photocurrent $I_{PC}$ for $V_{12} = 10$ V and $V_g = 50$ V as a function of the waveplate angle for different illumination angles. Dots indicate experimental data. Solid lines are fittings to equation 1. (b) Dependence of $C$ on the illumination angle, extracted from the fittings shown in (a). The dotted line is a parabolic fitting of the experimental data, shown as a guide to the eye.

## 5. CPC and illumination angle of incidence

Figure 4a shows the measured photocurrent for $\lambda = 790$ nm, $V_{12} = 10$ V and $V_g = 50$ V as a function of the waveplate angle for different illumination angles. From these measurements we extract the angle dependence of the helicity-dependent photocurrent, $C$, shown in Figure 4b. The dependence of CPC on the illumination angle allows us to extract information on the underlying physical mechanism.

As we discussed in ref. 11, CPGE cannot occur in a material with D$_{3h}$ symmetry, such as pristine 1L-MoSe$_2$, while CPDE can only give contributions proportional to $\sin(2\phi)$, which are odd upon inversion of the illumination angle $\phi$ and should cancel out for illumination normal to the crystal plane. For the device with direct metal / MoSe$_2$ contacts studied here we find that $C$ has both even and odd components upon inversion of the illumination angle. Furthermore, a nonzero helicity-dependent signal is observed even for normal-incidence illumination. This is in strong contrast with our results in h-BN encapsulated devices (see Supplementary Note 2 and ref 11). For these devices, a nonzero angle of incidence is needed to generate a measurable CPC. Simple symmetry arguments can be used to show that a nonzero CPC signal at $\phi = 0°$ can only appear if the symmetry of the crystal is reduced to at most a single mirror-plane symmetry.[11] Thus, our results establish that the presence of non-equivalent Schottky barriers in the vicinity of the metallic contacts results in a largely reduced symmetry of the electronic states, allowing for additional contributions to the CPC. In fact, as shown in Supplementary Note 3, we observe that the overall strength and angle dependence of the CPC varies from one set of contacts to another, further suggesting that this effect is largely affected by the local geometry near the electrodes.

## 6. Effect of the drain-source voltage on CPC

Finally, we evaluate the dependence of the CPC on the drain-source voltage. As mentioned above, for the sample without h-BN tunnel barriers a nonzero bias voltage needs to be applied in order to observe nonzero $C$ and $L$ photocurrent components. This is again in contrast with our results for h-BN encapsulated devices (see Supplementary Note 2 and ref. [11]), where a clear helicity-dependent photocurrent appears even in short-circuit configuration.

When $V_{12}$ is swept, both $C$ and $L$ increase with the absolute value of $V_{ds}$. However, the sign and amplitude of $C$ depend on the angle of incidence in a non-trivial way. We also observe that $C$ and $L$ are largely dependent on the selected set of source-drain contacts for a fixed angle of incidence (see Supplementary Note 3). Figure 5 shows the dependence of $C$ and $L$ on the drain-source voltage $V_{12}$ for two different angles of incidence ($\phi = +50°$ and $\phi = -50°$). For $+50°$ a nonzero helicity-dependent photocurrent $C$ is clearly observed for positive drain source voltages $V_{12}$, increasing monotonically with the applied voltage (see Fig. 5c). For negative voltages a smaller but measurable $C$ is observed. Intriguingly, the sign of $C$ is preserved when changing the sign of the drain-source voltage. However, when the illumination angle is inverted from $50°$ to $-50°$ the sign of $C$ flips from positive to negative, and a large signal is only observed for negative $V_{12}$ (Figure 5c). It is worth noting that the behavior described here is only valid for a specific set of contacts in the device and changes in a nontrivial way with the angle of incidence. $V_{ds}$-dependent measurements for additional contacts and illumination angles can be found in Supplementary Note 3. The fact that $C$ and $L$ modulate differently with the drain-source voltage for different sets of contacts is consistent with our hypothesis, *i.e.*, that the presence of nonhomogeneous Schottky barriers alters the symmetry of electronic states and thus enables additional contributions to the CPC, largely affecting the total measured CPC signal.

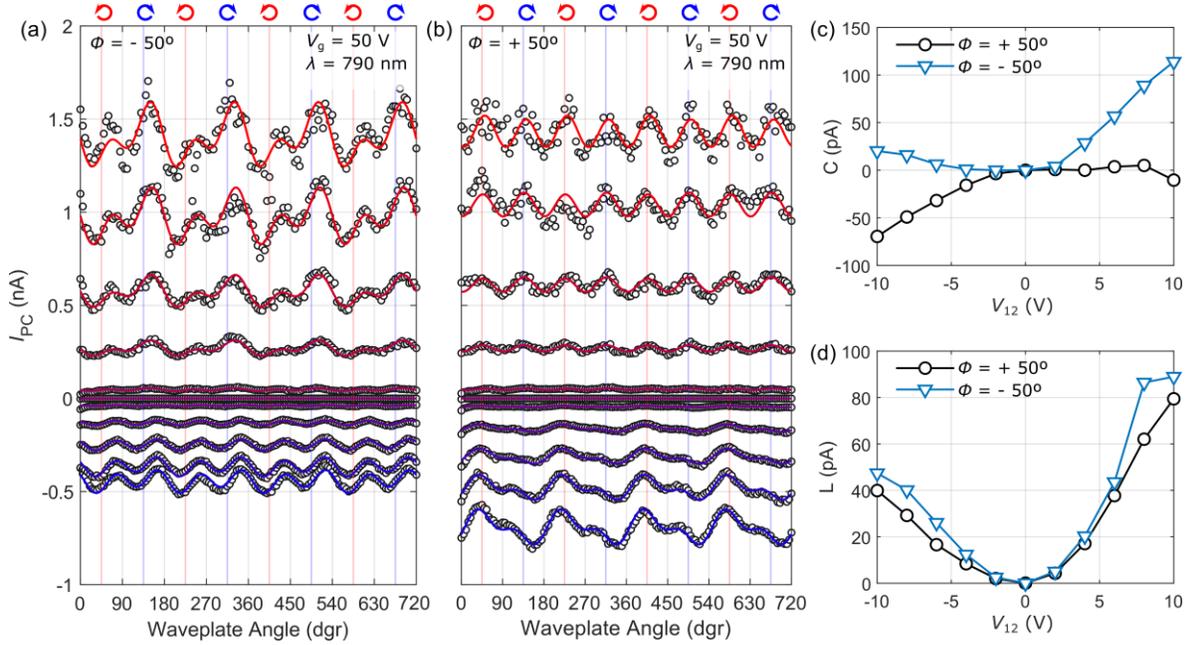

**Figure 5** – Dependence of C and L on the drain source voltage. (a-b) Photocurrent $I_{PC}$ as a function of the waveplate angle for different drain-source voltages $V_{12}$ between -10 V (blue) and +10V (red), for a light incidence angles $\phi = +50°$ (a) and $\phi = -50°$ (b). Dots indicate experimental data. Solid lines are fittings to equation 1. (c-d) Dependence of $C$ (c) and $L$ (d) on the drain source voltage, extracted from the fittings shown in (a) and (b).

## 7. Discussion

In a seminal theory work,[24] Moore and Orenstein showed that the presence of a nonzero Berry curvature in a 2D system can lead to the emergence of a CPGE that is maximum under illumination normal to the crystal plane, and modulates with the angle of incidence as $\cos(\phi)$. While earlier attempts have been made to detect this phenomenon in 2D TMDs, [10,11] the angular dependence observed there was not compatible with a Berry curvature-induced CPGE. In particular, a CPC was only observed for oblique illumination. In the light of our results, the device with direct metal-semiconductor yields a CPC that does not vanish for $\phi = 0°$ and contains contributions that are even upon inversion of the illumination angle. Thus, the symmetry of the observed CPC is compatible with the Berry curvature-induced CPGE. As discussed in earlier literature [11], a nonzero CPC for normal-incidence illumination, $\phi = 0°$, is only possible for a device with, at most, a single mirror-plane symmetry. Thus, we conclude that the presence of strong electric fields at the Schottky contacts reduces the symmetry of the MoSe$_2$ channel, enabling the observed CPC.

It is worth remarking that earlier measurements on helicity-dependent photoresponse carried out in 1L-TMD devices with direct metal-semiconductor contacts could also show contributions caused by a symmetry reduction near the contacts. For example, for the valley-Hall effect[16], N. Ubrig et al. recently showed that the helicity-dependent signal is strongly modified when the vicinity of the electrodes is exposed to light[13], which could be caused by a contact-induced symmetry breaking. In light of our measurements, inserting few-layer h-BN as tunnel barrier between the semiconductor channel and the metallic contacts minimizes possible effects of Schottky barriers on the device photoresponse, granting access to the intrinsic properties of a 2D-TMD.

Our results also show that CPCs can be very largely and nontrivially modulated by the illumination angle, even for incidence angles as small as 10°. However, most reports on helicity-dependent optoelectronic measurements rely on high numerical aperture objectives to focus the laser beam onto a small area of the sample. While this method has the advantage of granting micrometer spatial resolution, it comes at the price of losing resolution on the illumination angle, as the measured photoresponse will be averaged over a broad range of angles. Thus, in order to obtain a complete microscopic understanding of the helicity-dependent optoelectronic response of 2D-TMD devices, spatially-resolved experiments should be used in combination with angle-resolved measurements. We envision that the symmetry-breaking generated by Schottky contacts can also be used for engineering of CPC in 2D-TMD phototransistors, opening up possibilities to tuning the photoresponse for circularly polarized light at particular incident angles for angular-resolved photodetectors.

## Methods

*Fabrication of hBN-encapsulated devices* - We mechanically exfoliate atomically thin layers of MoSe$_2$ and h-BN from their bulk crystals on a SiO$_2$ (300 nm)/doped Si substrate. The monolayer MoSe$_2$ and bilayer h-BN are identified by their optical contrasts with respect to the substrate and their thickness is confirmed by atomic force microscopy. Using a polymer-based dry pick-up technique, we pick up the bilayer h-BN flake with a PC (poly(bisphenol A)carbonate) layer attached to a polydimethylsiloxane (PDMS) stamp. Then we use the same stamp to pick up the MoSe$_2$ flake directly in contact with the h-BN surface and we transfer the whole stack onto a bulk h-BN crystal, exfoliated on a different SiO$_2$/Si substrate. After the final transfer step, the PC layer is detached from the PDMS, remaining on top of the 2L-BN/MoSe$_2$/bulk-BN stack, and must be dissolved using chloroform. To further clean the stack, we anneal the sample in Ar/H$_2$ at 350 °C for 3 h.

*Electrode fabrication* – The same electrode fabrication process is followed for the two device geometries (with and without hBN tunnel barriers). First, we pattern the stacks by electron-beam lithography using PMMA (poly(methyl methacrylate)) as the e-beam resist. This is followed by e-beam evaporation of Ti(5 nm)/Au(75 nm) at $10^{-6}$ mbar and lift-off in acetone at 40 °C.

## Author Contributions

JQ and JH initiated the project. JH and TG performed the device fabrication. JQ and JH performed the measurements and analyzed the data, with MHDG assistance. BJvW, CHvW, and MHDG supervised the project. JQ wrote the manuscript with JH and MHDG assistance. All authors agreed on the final version of the manuscript.

## Acknowledgements

We acknowledge funding from the Dutch Foundation for Fundamental Research on Matter (FOM) as a part of the Netherlands Organization for Scientific Research (NWO), FLAG-ERA (15FLAG01- 2), the European Unions Horizon 2020 research and innovation programme under grant agreements No 696656 and 785219 (Graphene Flagship Core 2 and Core 3), NWO Start-Up (STU.019.014), NanoNed, the Zernike Institute for Advanced Materials, and the Spinoza Prize awarded to BJ van Wees by NWO. J.Q acknowledges financial support from the Agencia Estatal de Investigación of Spain (Grants MAT2016-75955, PID2019-106820RB-C22 and RTI2018-097180-B-100) and the Junta de Castilla y León (Grant SA256P18), including funding by ERDF/FEDER. MHDG acknowledges financial support from NWO Veni (15093).

**Competing Interests Statement**

The authors declare that there are no competing interests.

**Data Availability Statement**

The data that support the findings of this study are available from the corresponding author upon reasonable request.

Supplementary Material to: The role of device asymmetries and Schottky barriers on the helicity-dependent photoresponse of 2D phototransistors

*Jorge Quereda[1,2]\*, Jan Hidding[1], Talieh S. Ghiasi[1], Bart J. van Wees[1],*
*Caspar H. van der Wal[1] and Marcos H.D. Guimaraes[1].*

[1] Zernike Institute for Advanced Materials, University of Groningen, Groningen, The Netherlands.
[2] Nanotechnology Group, USAL-Nanolab, Universidad de Salamanca, Salamanca, Spain.

\* e-mail: j.quereda@usal.es

**Table of contents**

## Supplementary Note 1: Optical and AFM images of the device

In this section we focus on the fabrication of the non-encapsulated device. We address the reader to the methods section in the main text and the supplementary information of ref. 1 for details on the fabrication of the h-BN encapsulated device. Supplementary Figures 1a and b show an optical micrograph and an AFM image of the $MoSe_2$/h-BN heterostructure prior to contact fabrication. The $MoSe_2$ crystal presents monolayer and a three-layer regions and is cracked into 5 separate regions. The leftmost part of the crystal consists of a uniform monolayer. Supplementary Figure 2 shows an optical image of the final device after contact fabrication. Electrodes 1 to 4 are numbered accordingly with Figure 1a in the main text.

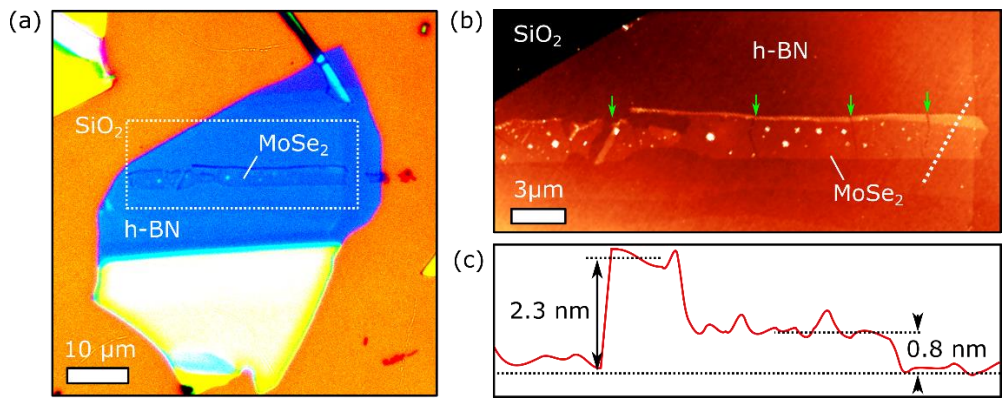

**Supplementary Figure 1** – (a) Enhanced-contrast optical microscope image of the $MoSe_2$ flake transferred on top of a few nm thick h-BN. (b) Tapping-mode AFM image of the region marked in (a). The $MoSe_2$ flake presents a series of cracks that split in in separate regions (indicated by green arrows in the figure). The leftmost region in the figure is a homogeneous monolayer, while the other four regions contain both monolayer and 3-layer $MoSe_2$ (c) Single AFM scan across the dashed white line indicated in (b), showing the monolayer (0.8 nm) and three-layer (2.3 nm) regions.

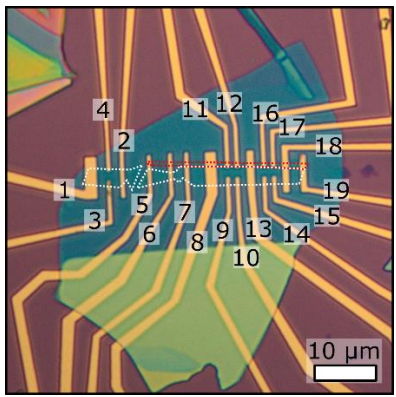

**Supplementary Figure 2** – Optical image of the non-encapsulated $MoSe_2$ device. The white and red dashed lines indicate the positions of the 1L and 3L-$MoSe_2$ flake regions, respectively. Contacts 1, 2, 3 and 4 are numbered accordingly with the schematic representation of Figure 1a in the main text.

**Supplementary Note 2: Additional CPC measurements in the h-BN encapsulated device**

Supplementary Figure 3 shows the helicity-dependent photocurrent measured in the hBN-encapsulated device with tunnel contacts as a function of the drain-source voltage, for $\lambda = 790$ nm and $\phi = +20°$. Differently from the device with direct metal/MoSe$_2$ contacts shown in the main text, here we observe a nonzero $C$ component at $V_{ds} = 0$ V, even when the polarization-independent component $I_0$ cancels out for this drain-source voltage.

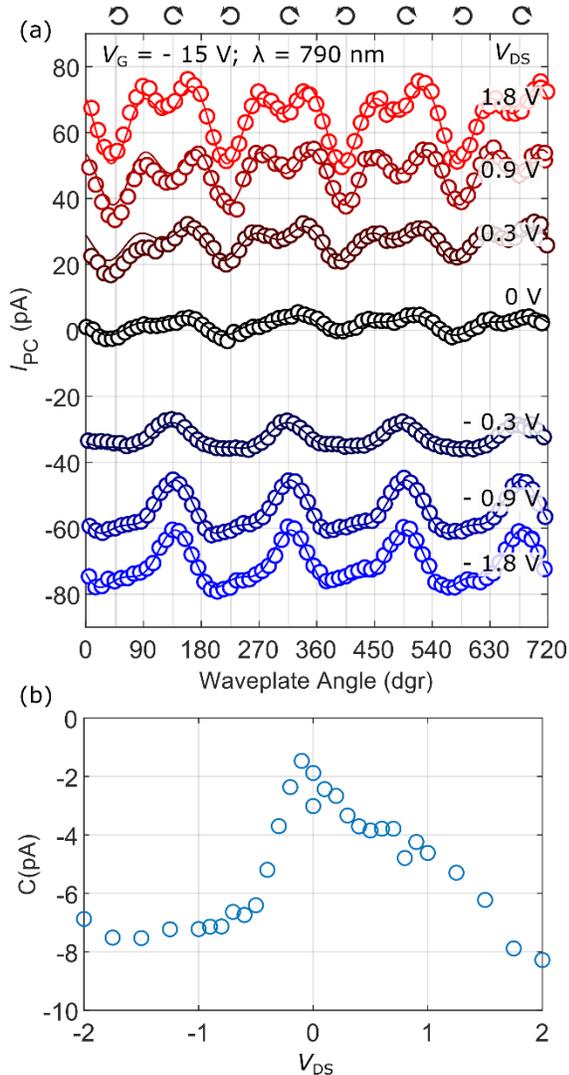

**Supplementary Figure 3** – Photocurrent $I_{PC}$ as a function of the waveplate angle for different drain-source voltages $V_{ds}$ between -1.8 V (blue) and + 1.8 V (red), for a light incidence angle $\phi = +20°$. Solid lines are fits to equation 1. (b) Dependence of $C$ on the drain-source voltage, extracted from the fittings shown in (a).

As shown in supplementary Figure 4, the helicity-dependent component of the photocurrent for the encapsulated sample is strongly dependent on the illumination angle $\phi$. In contrast with the non-encapsulated sample, here we observe that at normal incidence ($\phi = 0$) the helicity dependent component cancels out, and only a helicity-independent photocurrent is observed.

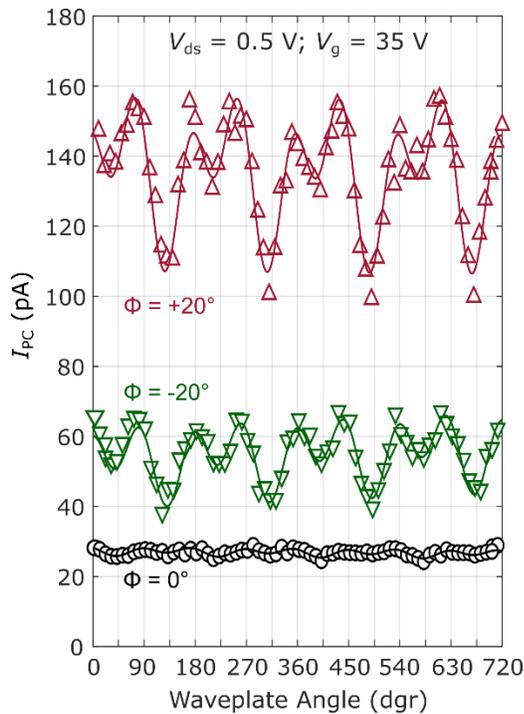

**Supplementary Figure 4** – Photocurrent $I_{PC}$ as a function of the waveplate angle for three different light incidence angles: $\phi = +20°$ (maroon upward triangles), $\phi = 0°$ (black circles) and $\phi = -20°$ (green downward triangles). Solid lines are fits to equation 1.

**Supplementary Note 3: Additional CPC measurements in the non-encapsulated device**

The results shown in the main text are obtained using two specific contacts (number 1 and 2). To examine the role of the contacts in these measurements, similar measurements were performed using another pair of contacts. In this section, the results obtained using contact 14 and 15 (see Supplementary Figure 2) are presented.

*Spectral measurements in different contacts*

As shown in Supplementary Figures 1 and 2, the device with direct metal-semiconductor contacts is not fully homogeneous, and also contains a three-layer part. Figure 5 shows the spectral dependence of the $I_0$, $C$ and $L$ components of $I_{PC}$ when using contacts 16 and 17 (see Supplementary Figure 2) as drain and source. For this situation the device response is mainly dominated by the three-layer region, and the spectral dependence of the $C$ parameter markedly changes with respect to the one shown in the main text.

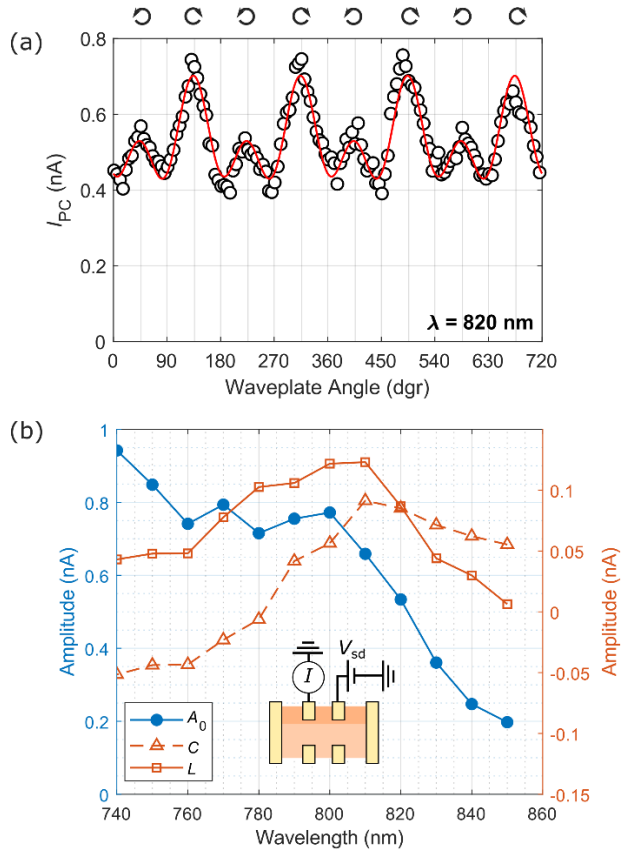

**Supplementary Figure 5** – Helicity-dependent spectra for contacts 16 and 17. (a) Photocurrent $I_{PC}$ as a function of the waveplate angle for $\lambda = 820$ nm, $V_{ds} = 5$ V and $V_g = 50$ V. The red line is a fitting to equation 1 in the main text (b) Spectral dependence of the $A_0$ (blue filled circles; left axis), $C$ (orange empty squares; right axis) and $L$ (orange empty triangles; right axis) parameters extracted from fittings to equation 1. Inset: Schematic drawing of the device and measurement geometry.

As shown in Supplementary Figure 6, when we repeat again the spectral measurement using a source contact placed at the three-layer region (contact 16) and a drain contact placed at the single-layer region (contact 14), we recover the spectral behavior described in the main text, *i.e.* the $C$ component becomes maximal at $\lambda = 790$ nm, on resonance with the A exciton transition, and drastically decreases for lower wavelengths, even when $I_0$ and $L$ remain large.

(a)

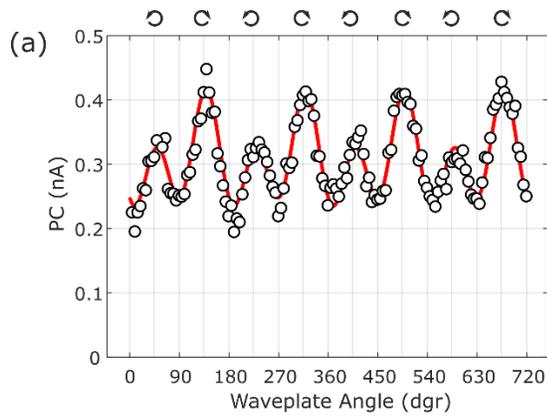

(b)

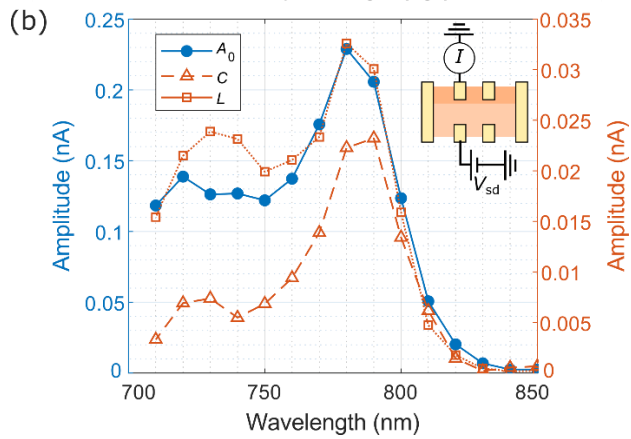

**Supplementary Figure 6** – Helicity-dependent spectra for contacts 16 and 14. (a) Photocurrent $I_{PC}$ as a function of the waveplate angle for $\lambda =$ 790 nm, $V_{ds}$ = 5 V and $V_g$ = 50 V. The red line is a fitting to equation 1 in the main text (b) Spectral dependence of the $A_0$ (blue filled circles; left axis), $C$ (orange empty squares; right axis) and $L$ (orange empty triangles; right axis) parameters extracted from fittings to equation 1. Inset: Schematic drawing of the device and measurement geometry.

*Dependence of CPC on the gate voltage*

Similar to section 4 in the main text, gate dependent measurements are performed. We apply gate voltages between $V_g = 0$ V and 50 V with steps of 5 V while keeping the drain source voltage constant ($V_{ds} = -10$ V). The sample is illuminated at $\lambda = 790$ nm at an incidence angle of $\phi = -50°$. The resulting photocurrent is monitored while rotating the quarter waveplate by two full rotations, as shown in Supplementary Figure 7. This allows us to more accurately remove drift effects. Subsequently, $I_0$, $C$ and $L$ are extracted by fitting the data to equation 1 in the main text.

In the main text we observe a monotonic increase of the $I_0$, $C$ and $L$-component with increasing gate voltage, where the ratio of the $L$ and $C$-component stay constant. Here we observe a similar gate voltage dependence. Again, the magnitude of all three components increases monotonically with the gate voltage. Note that the sign difference of the components compared to the main text is due to a 180° phase shift during the fitting. These results show that the gate voltage dependence of $C$ is similar for different contacts.

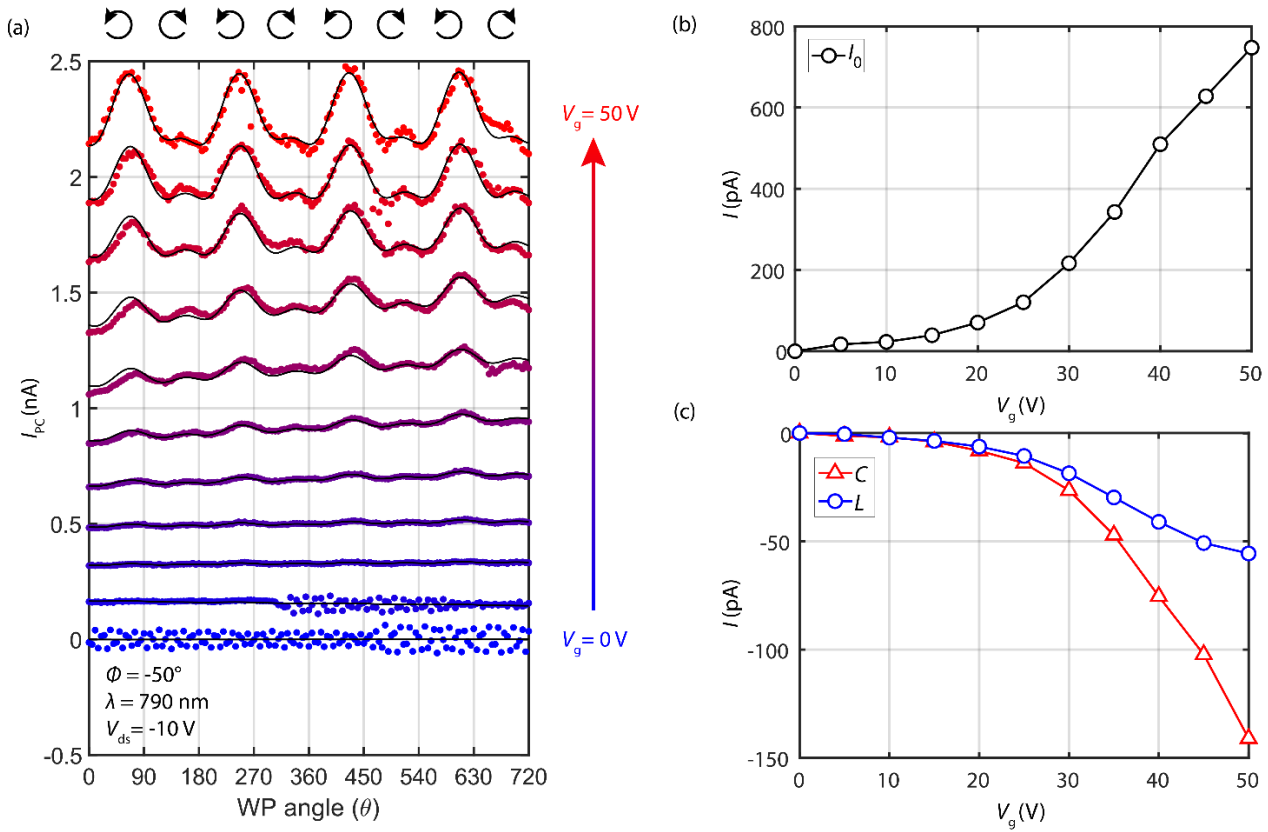

**Supplementary Figure 7** – Gate dependent photocurrent measurements at an incidence angle of -50° and a $V_{ds}$ of -10 V. The corresponding $I_0$, $C$- and $L$-component, obtained by fitting to equation 1, are plotted at the right.

The ratio between the *L*- and *C*-component however, does change upon switching to the different set of contacts. Contrary to the results depicted in the main text, we observe a change in the ratio of the *C* and *L*-components for different $V_g$ which indicates that the specific ratio of the *C* and *L* component is dependent on the set of contacts used.

**CPC and illumination angle**

Similar to section 5 in the main text, incidence angle dependent measurements were performed, ranging from $\phi = -60°$ to $60°$. Both the drain source voltage and the gate voltage are kept constant at $V_{ds}$ = -10 V and 50 V, respectively. As before, the sample is excited by a continuous wave laser with a wavelength of $\lambda = 790$ nm. The resulting photocurrent is measured while modulating the polarization of the light using a rotatable quarter waveplate. Supplementary Figure 8a shows the PC measurements obtained for the different incident angles. The extracted *C*-component from these measurements are depicted in Supplementary Figure 8b. As before, we observe a large dependence of *C* on the incidence angle with both even and odd components upon reversing the incidence angle.

However, comparing Supplementary Figure 8b to Figure 4b in the main text, we see distinct differences. In the main text, we observe a maximum *C*-component at and incidence angle of $\phi =$ 10°. Here, however, $C$ is maximized at $\phi = 30°$ and shows a larger magnitude compared to the $C$-component at $\phi = 10°$ in the main text. Similar those measurements, a small nonzero $C$-component is obtained at normal incidence ($\phi = 0°$).

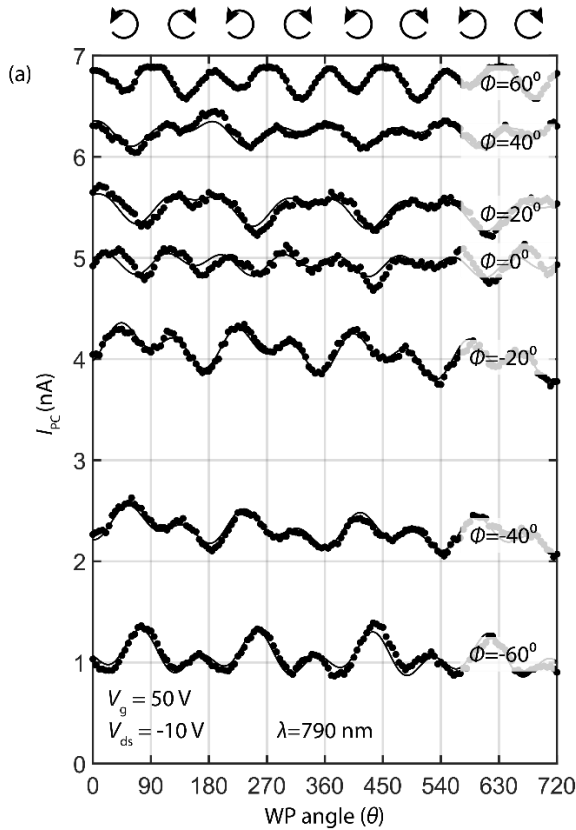

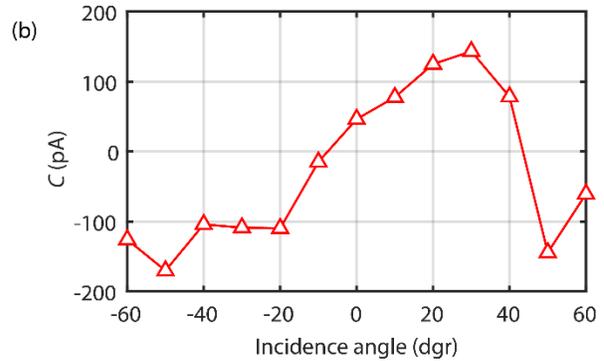

**Supplementary Figure 8** - Dependence of *C* on the incidence angle. (a) Photocurrent $I_{PC}$ for $V_{ds} = $ -10 V and $V_g$ = 50 V as a function of the waveplate angle for different incidence angles. Dots indicate experimental data. Solid lines are fittings to equation 1. (b) Dependence of *C* on the incidence angle, extracted from the fittings shown in (a).

The fact that we see distinct differences for different sets of contacts is in line with our assumption that nonhomogeneous Schottky barriers in the vicinity of the metallic contacts play a significant role in these measurements – as different contacts are likely to have different Schottky barriers, a different behavior is expected.

**Effect of the drain-source voltage on CPC**

Next, the drain-source voltage on the CPC is investigated. As before, we illuminate the sample at $\lambda = 790$ nm and apply a constant gate voltage of $V_g = 50$ V. We measure the photocurrent while rotating a quarter waveplate for different drain source voltages (ranging from -10 V to 10 V with steps of 2 V). Two opposite incidence angles of $\phi = -50°$ and $50°$, were chosen, as these angles showed a large CPC during the previous incidence-angle dependent measurement (Supplementary Figure 8b).

Supplementary Figure 9 shows the dependence of $C$ and $L$ on the drain-source voltage. The $C$-components obtained for $\phi = -50°$ and $\phi = 50°$ are comparable for both negative and positive drain-source voltages. This illustrates once more that a net nonzero $C$-component can be observed when i.e. large N.A. objectives are used to focus down the light. Compared to Figure 5 in the main text, the difference between the $C$-components of the opposite incidence angles is smaller. Furthermore, the $V_{ds}$-dependence on $C$ is significantly different compared to Figure 5 in the main text. There, a large positive $C$-component is obtained at positive $V_{ds}$ at negative incidence angles, while the measurements here show a $C$-component close to zero. Vice versa, at negative $V_{ds}$, a large and negative $C$-component is obtained here for negative incidence angles, while in the main text it is close to zero. At positive incidence angles, the differences are smaller. For this set of contacts, we observe a monotonic decrease of the $C$-component with decreasing $V_{ds}$ up to -10 V. This is in contrast to the $C$-component at $\phi = -50°$ shown in Figure 5.

This shows that, similar to the angle of incidence dependence, the CPC dependence on the $V_{ds}$ is largely influenced by the set of contacts used. Therefore, we conclude that the Schottky barriers at the contacts play an important role in both the magnitude and overall behavior of these signals. The $L$-components obtained for the opposite incidence angles overlap to a larger extent. However, where the $L$-component in the main text shows a symmetric response around $V_{ds} = 0$ V, an asymmetric response is obtained here. At positive $V_{ds}$, the $L$ component increases almost linearly, while at negative $V_{ds}$ a nonlinear decrease is observed for both incidence angles.

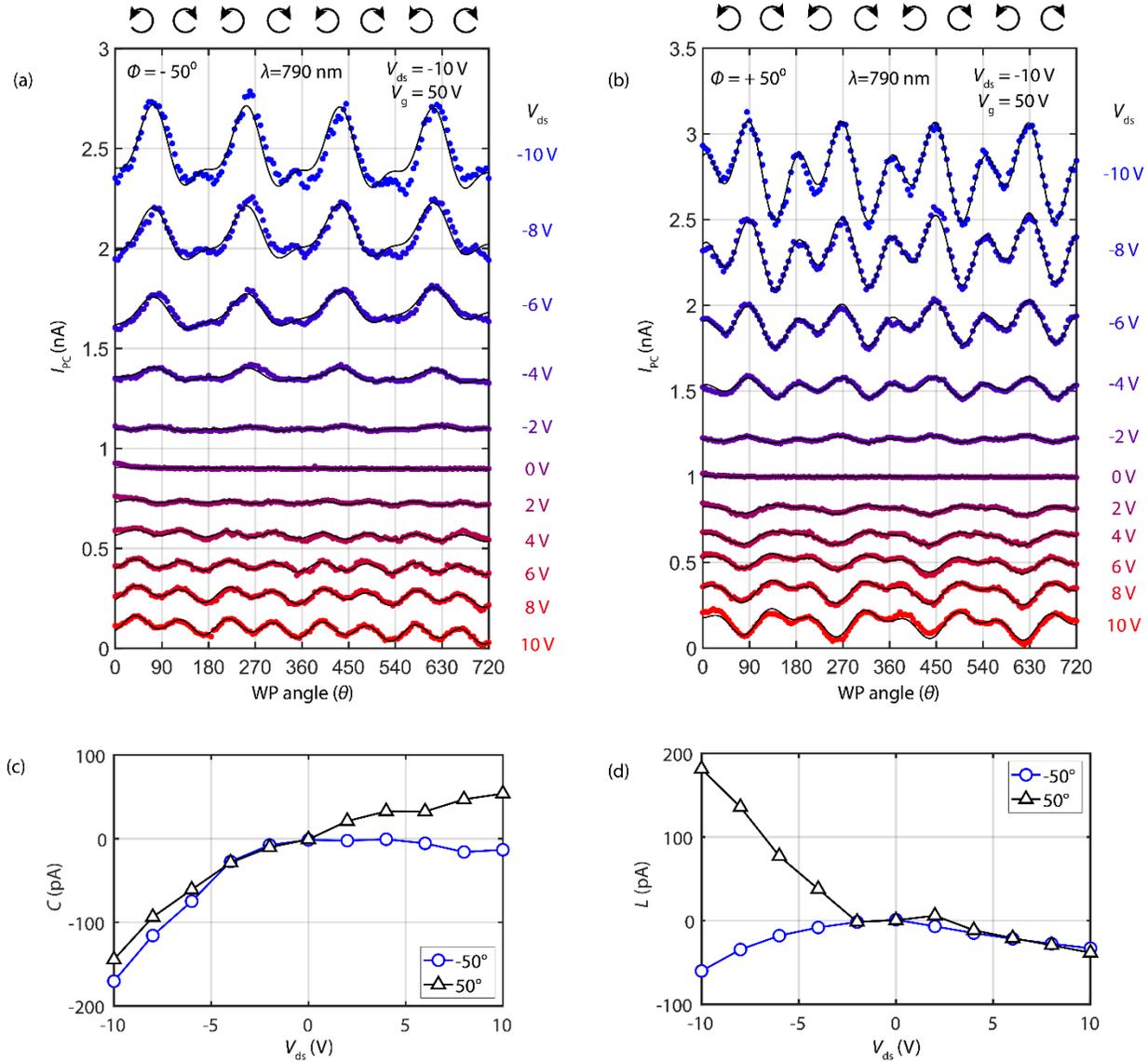

**Supplementary Figure 1** - Dependence of $C$ and $L$ on the drain source voltage. (a-b) Photocurrent $I_{PC}$ as a function of the waveplate angle for different drain-source voltages $V_{ds}$ between -10 V (blue) and +10 V (red), for a light incidence angles $\phi$=-50° (a) and $\phi$ =50° (b). Dots indicate experimental data. Solid lines are fittings to equation 1. (c-d) Dependence of $C$ (c) and $L$ (d) on the drain source voltage, extracted from the fittings shown in (a) and (b).